# Room-temperature Single Photon Emitters in Cubic Boron Nitride Nanocrystals

Gabriel I. López-Morales[1,2,3], Aziza Almanakly[1,4], Sitakanta Satapathy[1], Nicholas V. Proscia[1], Harishankar Jayakumar[1], Valery N. Khabashesku[5,6], Pulickel M. Ajayan[6], Carlos A. Meriles[1,7*], Vinod M. Menon[1,7*]

[1]Department of Physics, City College of the City University of New York, New York, NY 10031 USA
[2]Department of Chemistry, Lehman College of the City University of New York, Bronx, NY 10468, USA
[3]Ph. D. Program in Chemistry, The Graduate Center of the City University of New York, New York, NY 10016, USA
[4]Department of Electrical Engineering, The Cooper Union for the Advancement of Science and Art, New York, NY 10003, USA
[5]Baker Hughes, Center for Technology Innovation, Houston, TX 77040, USA
[6]Department of Materials Science and NanoEngineering, Rice University, Houston, TX 77005, USA
[7]Ph. D. Program in Physics, The Graduate Center of the City University of New York, New York, NY 10016, USA

*Email: cmeriles@ccny.cuny.edu, vmenon@ccny.cuny.edu

Color centers in wide bandgap semiconductors are attracting broad attention as platforms for quantum technologies relying on room-temperature single-photon emission (SPE), and for nanoscale metrology applications building on the centers' response to electric and magnetic fields. Here, we demonstrate room-temperature SPE from defects in cubic boron nitride (cBN) nanocrystals, which we unambiguously assign to the cubic phase using spectrally resolved Raman imaging. These isolated spots show photoluminescence (PL) spectra with zero-phonon lines (ZPLs) within the visible region (496–700 nm) when subject to sub-bandgap laser excitation. Second-order autocorrelation of the emitted photons reveals antibunching with $g^2(0) \sim 0.2$, and a decay constant of 2.75 ns that is further confirmed through fluorescence lifetime measurements. The results presented herein prove the existence of optically addressable isolated quantum emitters originating from defects in cBN, making this material an interesting platform for opto-electronic devices and quantum applications.

With an atomic structure equivalent to that of diamond, refractive index close to 2.1, an optical bandgap exceeding 10 eV, and the possibility of *n*- and *p*-type doping, cubic boron nitride (cBN) is attracting interest as a host for atomic defects with optical features across a wide region of the electromagnetic spectrum.[1-3] A combination of photoluminescence (PL) and electron paramagnetic resonance (EPR) studies have shed light on the optical and electronic properties of a range of point defects, from vacancies to heavy-metals and rare-earth elements.[4-9] Computational methods have also been used to study native and extrinsic defects in cBN, predicting the existence of an oxygen-related defects having isoelectronic properties to the well-known NV$^-$ center in diamond.[10-12] Thus far, however, experimental signatures from single atomic defects in this material have proven elusive. Mainly, this has been due to its hardness along with its synthesis requirements, which pose limitations on defect engineering methods; as well as the energy separation between the defect levels within the bandgap, often large enough to make them inaccessible through optical excitation. Cathodoluminescence (CL), which provides the necessary energies to reach deep intra-band defect levels in cBN has been successfully used in this context. However, electron excitation usually results in synchronized emission from nearby defect centers that can be excited by the same incoming electron, posing difficulties for single-defect studies and antibunching measurements which are key to unambiguously identify individual defects.[13, 14]

Here, we report on room-temperature single-photon emission (SPE) originating from atomic defects in cBN, some of which show spectral features like those previously reported for color centers in single-crystal cBN.[1, 4] We overcome some of the aforementioned limitations by using nanocrystalline cBN flakes, which have a high density of near-surface atomic defects accessible through optical laser excitation. By employing confocal Raman spectroscopy, we show that the optical regions of interest are indeed dominated by the cubic phase of BN, while small hexagonal boron nitride (hBN) quantities arising as a synthesis byproduct are found to phase-segregate in nm-sized regions.

cBN nanocrystals (dispersed in ethanol) obtained from PlasmaChem GmbH are drop-cast onto a Si/SiO$_2$ substrate, kept at 60 °C under ambient conditions. The as-prepared samples show a variable distribution of cBN nanocrystal sizes (Fig 1 (a)) that strongly depends upon drop-cast parameters, such as solution concentration and dispersion. Individual nanocrystals range in size (80–450 nm), but are usually found to be around 160 nm. These high-quality cBN crystals can be obtained either by high-pressure/high-temperature (HPHT) synthesis from boron- and nitrogen-containing precursors, or by HPHT treatments of hBN in the presence of catalysts.[15] As a result, there are usually small fractions of hBN present in cBN samples, especially in polycrystalline, powder-like and cBN nanocrystals.[15, 16] Thus, an important part of this study consist in characterizing possible phase-segregation of hBN within the nanocrystalline cBN, while ensuring that the optically-probed domains correspond to cBN-dominated regions.

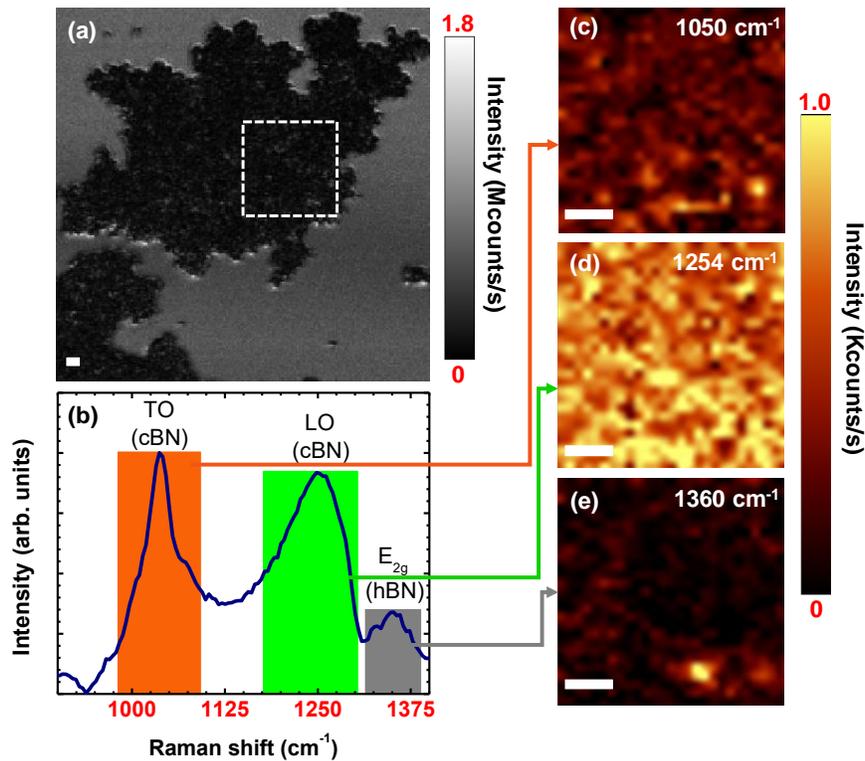

**Fig 1**. Confocal Raman spectroscopy of cBN nanocrystals. **(a)** Laser reflection imaging reveals large area coverage and conglomeration of cBN nanoflakes. **(b)** Raman spectrum collected by averaging over the squared region in **(a)** shows cBN LO and TO phonon modes strongly dominate over the E$_{2g}$ mode of hBN. **(c-e)** Confocal Raman spectroscopy for select phonon energies (shaded areas in **(b)**) show that the cubic phase of BN is dominant and uniform across the sample, while hBN regions seem to be localized possibly due to phase segregation. Scale bar = 2 μm.

Raman studies on BN show that cBN and hBN have spectrally separated vibrational signatures that can be used to identify and differentiate the two phases.[17-19] Specifically, both cBN and hBN have an optical phonon mode centered at Γ point which, due to ionic contributions, splits into longitudinal and transverse optical phonon modes (LO and TO) only in the case of cBN. These cBN modes are usually centered around 1304 and 1055 cm$^{-1}$, respectively, while the $E_{2g}$-symmetric phonon mode of hBN is centered around 1365 cm$^{-1}$.[17-19] Performing both optical and Raman confocal spectroscopy thus allows us to probe the phase profile of the spectral regions, while providing proof as to whether the collected fluorescence originates from defects in cBN nanocrystals or in residual hBN nanoflakes.

We perform Raman spectroscopy via a confocal Raman microscope (alpha300 R; WITec GmbH), equipped with a 100x (1.0NA) microscope objective and an SHG Nd:YAG laser (532 nm). Collected photons are sent to a spectrometer with a CCD-camera (1024×128 pixel, Peltier cooled to 65°C) rejecting reflected and elastically scattered photons by using an edge filter. The nominal spectral resolution is found to be ~5 cm$^{-1}$ per CCD pixel. Due to low scattering cross sections, Raman signatures from both BN phases tend to be close to the noise level, especially when dealing with nanocrystalline samples. A well-resolved Raman spectrum from the different BN phases is thus obtained by averaging over specific scanned areas (dashed square in Fig 1 (a)) while integrating for 0.5 s and illuminating at ~12.0 mW laser power (Fig 1 (b)). The main Raman spectral features extracted from this and several other locations include two dominant peaks centered at ~1050 and ~1254 cm$^{-1}$, which arise from TO and LO modes in cBN, respectively, and a third, lower intensity peak centered at ~1360 cm$^{-1}$ arising from the $E_{2g}$ phonon mode in hBN. For confocal Raman imaging, the 50 μm core of a multimode fiber serves as the pinhole, which leads to a focal depth of ~1 μm, while the diffraction-limited spot results in a lateral resolution of

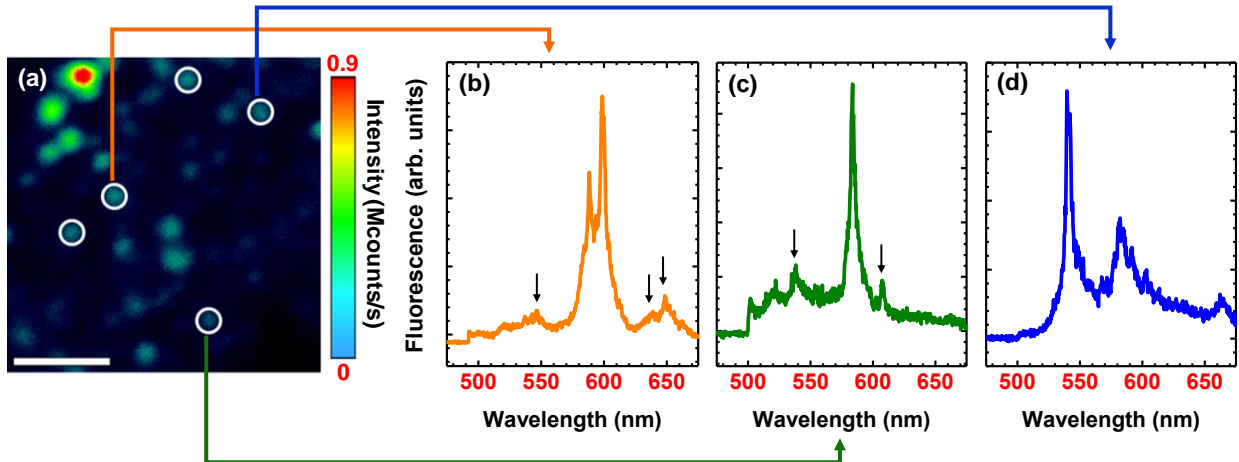

**Fig 2**. Steady-state confocal spectroscopy of cBN nanoflakes. **(a)** Confocal imaging of a representative cBN region reveals an array of isolated bright spots. White circles denote spots that showed sharp and spectrally separated ZPLs. **(b-d)** Representative optical spectra for different bright spots identified with (orange, green and blue) arrows in **(a)**. While **(b, c)** show spectral features comparable to those reported for defects in cBN (black arrows), **(d)** reveals a phonon replica matching the $E_{2g}$ phonon energy in hBN demonstrating that both cBN and hBN defects are present across the sample. Scale bar denotes 2 μm.

~1–2 $\mu$m. Scanning over the same region while only collecting one of the specific phonon energies in Fig 1 (b) reveals an almost uniform distribution for both TO and LO cBN modes (Fig 1 (c, d) respectively), while showing a non-uniform distribution for the $E_{2g}$ hBN mode (Fig 1 (e)). The small amount of residual hBN is thus observed to be localized at specific spots, mainly resulting from phase segregation. Further, the ratio, $\frac{I(E_{2g})}{I(TO+LO)} = 0.0615 \pm 0.0053$ averaged over multiple regions, confirms that these are indeed cBN-dominated nanocrystalline domains. We observe variable linewidths and small energy shifts in all the BN phonon modes throughout different sample locations, possibly due to size effects, local strain and relaxation of both phases to nanocrystalline states.[17, 19] However, all Raman measurements demonstrate that the cubic phase of BN dominates across the sample, while small hBN quantities are found to segregate within nm-sized regions.

To elucidate the optoelectronic properties of defect-related color centers in cBN nanocrystals, room-temperature confocal spectroscopy is performed via a home-built confocal microscope, equipped with a 50x (0.83 NA) objective and a 460 nm continuous-wave (CW) laser (L462P1400MM, ThorLabs). The collected fluorescence is filtered from the excitation source by using a 500 nm long-pass filter (LPF) and sent through a fiber beam splitter allowing for real-time spectral analysis and time-resolved measurements. The excitation source, with spot-size of about ~ 1 µm, is kept at an average power below 200 µW to avoid/suppress blinking and bleaching of single emitters. Despite the inhomogeneity in the spatial distribution of flakes owing to the drop-casting technique, a collection of isolated bright spots across large sample areas is systematically observed upon sub-bandgap excitation (Fig 2 (a)), indicating the presence of defect centers within the cBN nanoflakes independent of the sample preparation process. Most of the isolated bright spots inside the cBN nanoflakes usually give broad fluorescence features. Usually, two main cases are observed: (1) real-time spectral analysis shows substantial blinking of different zero-phonon lines (ZPLs) associated with different centers inside the same diffraction-limited spot (defect ensembles); (2) individual ZPLs arise on top of a broadband background, which is present in the nearby nanoflake regions. The majority of the ZPLs arising from single defects, however, are found to share similar spectral response, as can be observed by comparing Fig 2 (b) and Fig 2 (c), some of which resemble features previously reported from color centers in cBN, e.g., under UV-excitation.[4] Other ZPLs tend to resemble those observed in hBN defects (Fig 2 (d)), e.g., with phonon replicas matching the $E_{2g}$ phonon mode energy (~165 meV). Thus, it is noted that both cBN and hBN defects are found across the sample. However, the majority of the observed emitters are unlikely to originate from hBN defects due to: (1) their uniform distribution across the sample and the high extent of hBN localization (Fig 1); (2) mismatch between the obtained spectral features and those reported for hBN defects; (3) the similarities between most of the observed emitters and those previously reported in cBN.

To further study the dynamics and photon correlation of defects in cBN nanoflakes, we turn to room-temperature time-resolved and correlation spectroscopy. CW photon correlation measurements are performed via Hanbury-Brown-Twiss (HBT) interferometry (Picoquant, Picoharp 300) using two avalanche photodiode detectors (APD) and 460 nm laser excitation. On the other hand, lifetime measurements are performed using the 488 nm line of an 80 MHz, 500-fs pulsed tunable laser (Toptica FemtoFiber TVIS). Results from these studies are summarized in Fig 3. The obtained spectrum from a single defect center in cBN is shown in Fig 3 (a), with features indicated by arrows matching the shape and distribution of those previously reported.[4] In our study, however, these features appeared much more separated (~130 meV), while preserving equidistance from the main peak. Given the similarities between these shifts and the phonon energies for cBN (~130 and ~160 meV), they are likely to be phonon-mediated replicas of the main peak, hinting at a unique electronic defect structure with electron-phonon coupling that may be characteristic of defects in cBN. The fluorescence time-trace of this single defect center under continuous excitation showed substantial blinking (Fig 3 (b)), a signature characteristic of isolated quantum emitters, likely due to different charge-states of the excited defect. Photon correlation of the collected fluorescence through the second-order autocorrelation function, $g^2(\tau)$, gives rise to a clear dip at zero-time delay (Fig 3 (c)), with $g^2(0) = 0.229$ proving the quantum nature of the emitter. A slight bunching component with time constant of ~30 ns is observed, which points at a possible metastable and/or charge state that may correlate to the blinking behavior observed during fluorescence time traces.[5, 13, 20] Finally, lifetime measurements reveal an exponential decay (Fig 3 (d)) from which a radiative time constant of 2.753 ns is obtained. This exponential decay is further confirmed directly from CW $g^2(\tau)$ measurements and falls well within the range of fluorescence lifetimes previously reported from point defects in cBN.[1, 5] Exposure to external magnetic fields of up to 0.1 T and/or circularly polarized excitation yield no discernible change in the fluorescence, hence suggesting magneto-optically-active defects — including the $O_NV_B$ center, predicted to have a response similar to that of the NV⁻ in diamond — may not be widespread. We warn, however, most emitters tend to fade upon prolonged exposure to the laser beam, hence indicating additional work — e.g., exploiting time-modulated magnetic fields — will be required to expose weak magneto-optical contrast.

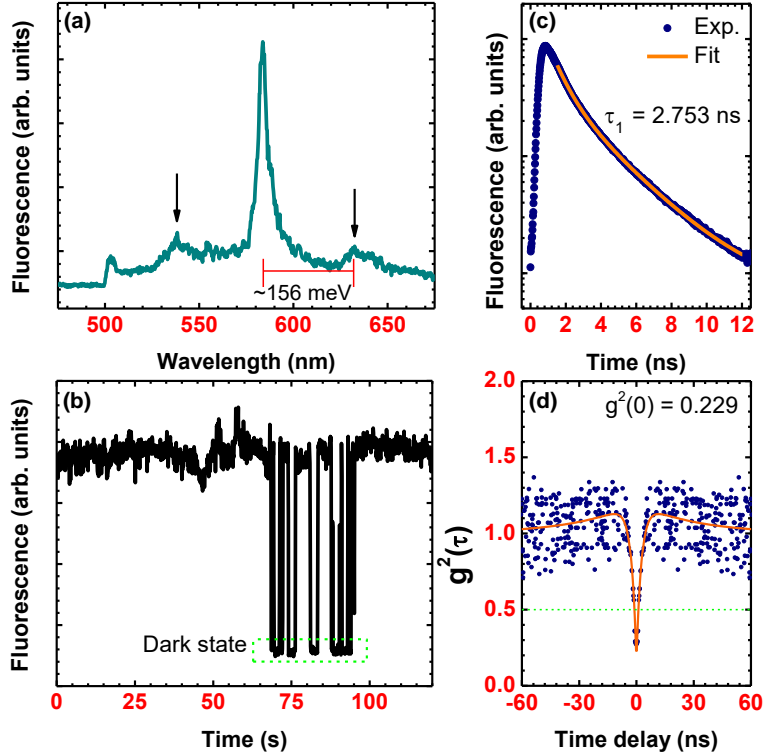

**Fig 3**. Time-resolved fluorescence spectroscopy of a single defect center in cBN. **(a)** Optical spectrum of an isolated color center under 460 nm continuous-wave (CW) excitation. Blue arrows indicate features that match those previously observed for the GB-1 color center under UV-excitation. Inset shows a zoomed-in confocal scan of this emitter (dotted circle). **(b)** Time-trace of the collected fluorescence showing fast transitions to a dark-state, characteristic of most of single-emitters. **(c)** Second-order autocorrelation function, showing photon antibunching at zero-time delay with $g^2(0) = 0.258$. **(d)** Fluorescence lifetime under 488 nm pulsed excitation. Solid line represents a biexponential decay fit to the experimental data (orange circles). Scale bar denotes 1 $\mu$m.

In summary, we have demonstrated room-temperature SPE from defects in cBN nanoflakes. Confocal images show isolated bright spots across the nanoflakes, while Raman spectroscopy confirms that these are indeed cBN-dominated nanocrystalline domains with the ratio, $I(E_{2g})/I(TO+LO)$ ~ 0.06. The amount of residual hBN is localized at different spots (phase segregation) and further confirms the cBN-origin of the observed color centers. Fluorescence spectroscopy reveals numerous ZPLs in the visible region, mainly covering 496–700 nm. Interestingly, the optical signatures of some of the studied defects resemble those previously reported[4] for color centers in cBN, with a clear shift (~130 meV) on the separation between the main peak and its low-intensity features. These features hint at an electronic structure unique to these color centers, while suggesting electron-phonon coupling similar to that of defects in cBN, e.g., the GB-1 centers.[1,4] These results open interesting possibilities in defect engineering based on cBN, combining properties similar to its hexagonal counterpart while providing a wider bandgap, ease of being ion-bombarded and a new group of diverse defect-related color centers.

**Acknowledgements:**
V.M.M. and C.A.M. acknowledge support from the National Science Foundation through grants NSF-1906096 and NSF-1726573. C.A.M. also acknowledges support from Research Corporation for Science Advancement through a FRED Award; all authors acknowledge partial funding and access to the facilities and research infrastructure of the NSF CREST IDEALS, grant number NSF-HRD-1547830. VM acknowledges useful discussions with Prof. Igor Aharonovich.